\documentclass[a4paper]{jpconf}
\usepackage{epsfig}
\usepackage{graphicx}
\usepackage{dcolumn}
\usepackage{bm}
\usepackage{amsmath, amsthm, amssymb}

\newcommand{\bea}{\begin{eqnarray}}
\newcommand{\eea}{\end{eqnarray}}

\begin{document}
\title{Membrane Paradigm and Holographic Hydrodynamics}

\author{Christopher Eling$^1$, Yasha Neiman$^2$, Yaron Oz$^2$}

\address{$^1$ SISSA, Via Bonomea 265, 34136 Trieste, Italy and INFN Sezione di Trieste, Via Valerio 2, 34127 Trieste, Italy}
\address{$^2$ Raymond and Beverly Sackler School of
Physics and Astronomy, Tel-Aviv University, Tel-Aviv 69978, Israel}

\ead{cteling@sissa.it, yashula@gmail.com, yaronoz@post.tau.ac.il}

\begin{abstract}

We discuss recent work showing that in certain cases the membrane paradigm equations governing the dynamics of black hole horizons can be recast as relativistic conservation law equations. In the context of gauge/gravity dualities, these equations are interpreted as defining the viscous hydrodynamics of a holographically dual relativistic field theory. Using this approach, one can derive the viscous transport coefficients and the form of the entropy current for field theories dual to gravity plus matter fields.

\end{abstract}

\section{Introduction}

In the past decade there has been much interest in the hydrodynamics of relativistic conformal field theories (CFTs), largely due to the AdS/CFT correspondence between (quantum) gravitational theories on asymptotically Anti-de-Sitter (AdS) spacetimes in $(d+1)$ dimensions and CFTs in $d$ dimensions \cite{adscft}.
In \cite{Bhattacharyya:2008jc} it was shown explicitly that the $d$-dimensional CFT hydrodynamics equations are dual to the gravitational field equations describing the evolution of large scale perturbations of the $(d+1)$-dimensional black brane. This has been dubbed the fluid-gravity correspondence. It has been noticed that this correspondence naturally completes the picture of the old membrane paradigm \cite{Damour,membrane} of the 1980's, where black hole horizons are thought of as being analogous to fictitious viscous fluids. In this case, however, the dynamics of the entire spacetime is encoded holographically into the fluid living at the boundary of the spacetime and not at the event horizon \cite{Hubeny:2009zz}.

On the other hand, at least in some circumstances, the horizon membrane paradigm can be modernized within the context of hydrodynamics as an effective field theory and the fluid-gravity correspondence. The starting point is an equilibrium solution containing a timelike Killing vector field and a stationary causal horizon. This solution is associated with a thermal state at uniform temperature. A general thermal state out of equilibrium is described by an inhomogeneous black hole, which is not an exact solution. However, when a hydrodynamic limit exists, we can work in a derivative expansion, assuming that there is no singularity at the horizon. In \cite{Eling:2010hu} we applied this approach to the event horizon of a boosted Einstein-Yang-Mills black brane in an asymptotically AdS spacetime. At lowest orders in derivatives, the set of Einstein and Maxwell equations projected into the horizon surface  describe the hydrodynamics of a CFT with anomalous global non-abelian charges. Here we will briefly describe the formalism used in \cite{Eling:2010hu} and show how it can be applied to the simplest case of an uncharged black brane \cite{Eling:2009sj}.

\section{Horizon Geometry}

In the bulk spacetime, we choose coordinates of the form $x^A = (r, x^\mu)$. The $x^\mu$ are coordinates on the horizon $\mathcal{H}$.  $r$ is a transverse coordinate, with $r = 0$ on the horizon. $\partial_A r$ is a null covector tangent to the $\mathcal{H}$. When raised with the metric, it gives a vector field $\ell^A = g^{AB}\partial_B r$ which is both normal and tangent to the horizon, and tangent to its null generators. In components, we have $\ell^A = (0, \ell^\mu)$.

The pullback of $g_{AB}$ into $\mathcal{H}$ is the degenerate horizon metric $\gamma_{\mu\nu}$. Its null directions are the generating light-rays of $\mathcal{H}$, i.e. $\gamma_{\mu\nu}\ell^\nu = 0$. The Lie derivative of $\gamma_{\mu\nu}$ along $\ell^\mu$ gives us the shear/expansion tensor, or ``second fundamental form'':
\begin{align}
	\theta_{\mu\nu} \equiv \frac{1}{2}\mathcal{L}_{\ell}\gamma_{\mu\nu}. \label{eq:theta_def}
\end{align}
We can write a decomposition of $\theta_{\mu\nu}$ into a shear tensor $\sigma^{(H)}_{\mu\nu}$ and an expansion coefficient $\theta$:
\begin{align}
	\theta_{\mu\nu} = \sigma^{(H)}_{\mu\nu} + \frac{1}{d-1}\theta\gamma_{\mu\nu}. \label{eq:theta_decompose}
\end{align}
We found it is convenient raise indices with $(G^{-1})^{\mu\nu}$, which is the inverse of any matrix $G_{\mu\nu}$ of the form
\begin{align}
	G_{\mu\nu} = \lambda\gamma_{\mu\nu} - b_\mu b_\nu; \quad b_\mu\ell^\mu \neq 0. \label{eq:G}
\end{align}
Here we introduced the superfluous scalar field $\lambda$ for later convenience: it will turn out that a matrix of the form \eqref{eq:G} coincides at leading order with the metric $h_{\mu \nu} = \eta_{\mu\nu}$ associated with the hydrodynamic dual.

Since $\gamma_{\mu\nu}$ is degenerate, one cannot use it to define an intrinsic connection on the null horizon, as could be done for spacelike or timelike hypersurfaces. The bulk spacetime's connection does induce a notion of parallel transport in $\mathcal{H}$, but only along its null generators. This structure is not fully captured by $\gamma_{\mu\nu}$; instead, it is encoded by the extrinsic curvature, or 'Weingarten map' $\Theta_\mu{}^\nu$, which is the horizon restriction of $\nabla_A\ell^B$:
\begin{align}
	\Theta_\mu{}^\nu = \nabla_\mu\ell^\nu.
\end{align}
%
One can show that given an arbitrary $G_{\mu\nu}$ of the form \eqref{eq:G}, $\Theta_\mu{}^\nu$ can be written as:
\begin{align}
	\Theta_\mu{}^\nu = \lambda\theta_{\mu\rho}(G^{-1})^{\rho\nu} + c_\mu\ell^\nu; \quad c_\mu\ell^\mu = \kappa. \label{eq:Theta}
\end{align}
$\kappa$ is the ``surface gravity",  which measures the non-affinity of $\ell^\mu$. The covector $c_\mu$ encodes the degrees of freedom in $\Theta_\mu{}^\nu$ which are independent of $\gamma_{\mu\nu}$.  In the hydrodynamics, these degrees of freedom will roughly correspond to the velocity and temperature fields.

In their usual form, the Gauss-Codazzi equations involve the divergence of a hypersurface's extrinsic curvature. Our null horizon, however, does not possess an intrinsic connection, and a covariant divergence of $\Theta_\mu{}^\nu$ cannot be defined. The solution is to define a tensor \emph{density} constructed out of $\Theta_\mu{}^\nu$ in the following manner \cite{Jezierski:2001qv}:
\begin{align}
	Q_\mu{}^\nu = v(\Theta_\mu{}^\nu - \kappa\delta_\mu^\nu),
\end{align}
where $v$ is a scalar density equal to the horizon area density.  It turns out that the horizon-intrinsic covariant divergence $\bar{\nabla}_\nu Q_\mu{}^\nu$ is uniquely defined.
%
%
Using this divergence, the null Gauss-Codazzi equation can be written as
%
\begin{align}
	R_{\mu\nu}S^\nu = D^{(G)}_\nu\left(\lambda v\theta_{\mu\rho}(G^{-1})^{\rho\nu}\right) + v\theta\partial_\mu\ln\sqrt{\lambda}
		+ c_\mu \partial_\nu S^\nu + 2S^\nu\partial_{[\nu}c_{\mu]} - v\partial_\mu\theta, \label{eq:GC_basic}
\end{align}
where we have defined the area entropy current $S^\mu = v \ell^\mu$. This is the form we we will use for calculations.

\section{AdS Black Brane and Ideal Hydrodynamics}

The vacuum Einstein equations with negative cosmological constant can be expressed as
\begin{align}
R_{AB}+d g_{AB} = 0.
\end{align}
These equations have the homogeneous (boosted) black brane solution
\begin{align}
ds_{(0)}^2 = -2 \ell_\mu dx^\mu dr - (r+R)^2 f \ell_\mu \ell_\nu dx^\mu dx^\nu + (r+R)^2 P_{\mu \nu} dx^\mu dx^\nu, \label{eq:BBmetric}
\end{align}
where $f = 1-(\frac{R}{r})^d$. $\ell^\mu$ is normalized $\eta_{\mu \nu} \ell^\mu \ell^\nu = -1$ and $P_{\mu \nu} = \eta_{\mu \nu} + \ell_\mu \ell_\nu$. The horizon quantities defined in the above section can be expressed in terms of the Bekenstein-Hawking entropy density $s= v^{(0)}/4= R^{d-1}/4$ and the Hawking temperature $T= \kappa^{(0)}/2\pi=(4/\pi d) R$
\begin{align}
 S^{(0)\mu} &= 4 s\ell^\mu; & \Theta_\mu^{(0)\nu} &= -2\pi T \ell_\mu \ell^\nu; & \gamma^{(0)}_{\mu\nu} &= (4s)^{\frac{2}{d-1}} P_{\mu\nu}; & c^{(0)}_\mu &= -2\pi T\ell_\mu, \label{eq:Zeroth}
\end{align}
where the superscript zero anticipates inhomogeneous corrections.  
%
%
Consider again the configuration \eqref{eq:BBmetric}, but with $(\ell^\mu,s)$ slowly varying functions of $x^\mu$ rather than constants. We will interpret this $x^\mu$-dependence by treating these quantities as fields on the brane's horizon $\mathcal{H}$. Various quantities and equations can be expanded order by order in powers of the small $\partial_\mu$ derivatives. We use the symbolic small parameter $\varepsilon$ to count these powers. We will refer to the power of $\varepsilon$ involved as the ``order'' of a quantity or an equation. In general, \eqref{eq:BBmetric} with $x^\mu$-dependent parameters will not be a solution of the field equations.

Our analysis will take place on the event horizon $\mathcal{H}$ of the exact inhomogeneous solution. In the configuration \eqref{eq:BBmetric} with inhomogeneous $(\ell^\mu,s)$, the hypersurface $r=0$ is null, and doesn't intersect the singularity. It is therefore an event horizon for the zeroth-order solution, and a zeroth-order approximation for the horizon of the corresponding exact solution. For the corrected solution, we will still use coordinates so that $r = 0$ on $\mathcal{H}$.

The Gauss-Codazzi equation \eqref{eq:GC_basic} reads at first order,
\begin{align}
	c^{(0)}_\mu \partial_\nu S^{(0)\nu} + 2S^{(0)\nu}\partial_{[\nu}c^{(0)}_{\mu]} =  O(\varepsilon^2). \label{eq:GC_1}
\end{align}
Plugging in the zeroth-order values,
%
and using the thermodynamic identities $\epsilon + p = Ts$ and $dp = sdT$, we rearrange the equation to get the conservation of a perfect fluid stress tensor
\begin{align}
\partial_\nu T^\nu_\mu = \partial_\nu ((\epsilon+p) \ell_\mu \ell^\nu + p \delta^\nu_\mu) = 0.
\end{align}
For the black brane one can show that $\epsilon = (d-1) p$, so the fluid is conformal, as expected.

\section{Viscous Hydrodynamics}

Now we want to consider the Gauss-Codazzi equations at second order. Obviously, before we write down these equations, we must consider the possible first-order corrections to the horizon ansatz \eqref{eq:Zeroth}. First of all, at viscous order in hydrodynamics there is naturally an ambiguity in definition of the fluid four-velocity. We make the (somewhat non-standard, but here natural) choice that the four-velocity is the direction of the entropy velocity; that is we fix the direction of the entropy velocity $\ell^\mu$ and the magnitude of the entropy density $s$ by requiring that they have the values in eq. \eqref{eq:Zeroth} without corrections.
%
%
Due to this condition, the correction to $\gamma_{\mu\nu}$ must be transverse to $\ell^\mu$ and traceless with respect to $\eta_{\mu\nu}$. We will find that the precise form of $\gamma^{(1)}_{\mu\nu}$ is otherwise irrelevant to the constraint equations.

Let us now turn to the $c_\mu$ component of the extrinsic curvature. For our uncharged black brane, the most general correction to $c_\mu$ reads
\begin{align}
  c_\mu^{(1)} = \mathcal{A} \ell_\mu \partial_\nu\ell^\nu + \mathcal{B} \partial_\mu s, \label{eq:c_1}
\end{align}
where $\mathcal{A}$ and $\mathcal{B}$ are functions of $s$.

The requirement for the constraint equations to take the form of conservation laws will place restrictions on these functions. To start, we note that the horizon's first-order shear/expansion tensor can be derived directly from the zeroth-order metric via \eqref{eq:theta_def}. Using the ideal equations, one finds that $\theta^{(1)}=0$ and $\theta_{\mu \nu}= (4s)^{\frac{2}{d-1}} \pi_{\mu \nu}$. where $\pi_{\mu \nu}$ is the fluid shear tensor $P^\rho_\mu P^\sigma_\nu \partial_{(\rho}u_{\sigma)} - \frac{1}{d-1}P_{\mu\nu} \partial_\rho u^\rho$.  Let us now evaluate the second order pieces of the Gauss-Codazzi equation \eqref{eq:GC_basic}.
%
Plugging in \eqref{eq:c_1}, and using the ideal order hydrodynamics equations we find
\begin{align}
\partial_\nu \left(4 s \pi_\mu^\nu\right) + 4\partial_\nu \left(\mathcal{A} s P^\nu_\mu (\partial_\rho \ell^\rho)\right) - 4\mathcal{A} (\partial_\rho \ell^\rho) \partial_\mu s.
\end{align}
The first two terms can contribute to a relativistic conservation law equation. However, the last does not and therefore we must set $\mathcal{A}=0$. As a result, at second-order, the Gauss-Codazzi equations have the form
\begin{align}
\partial_\nu (T_\mu^{(0)\nu}+T_\mu^{(1)\nu}) = \partial_\nu\left((\epsilon+p) \ell_\mu \ell^\nu + p \delta^\nu_\mu-\frac{s}{2\pi}\pi_\mu^\nu \right) = O(\varepsilon^3),
\end{align}
which are the relativistic hydrodynamics equations for a conformal fluid with shear viscosity $\eta/s=1/4\pi$.

\section{Discussion}

It is remarkable that in our analysis the dynamics of the null horizon does not encode an arbitrary hydrodynamic system: once the choices discussed above are made, the transport coefficients are uniquely determined. Specifically, the shear viscosity, the bulk viscosity (and in general, other quantities like conductivity, see \cite{Eling:2010hu}) are fixed by the null hypersurface equations to specific functions of state, while for a general fluid they may be arbitrary (under the restriction of positive semi-definiteness).

Note however that in relation to membrane paradigm picture of \cite{Hubeny:2009zz}, our construction cannot capture all the dynamics of the theory at higher orders in the hydrodynamic $\varepsilon$ expansion. Nevertheless, to viscous order we are able to obtain the hydrodynamics of the boundary fluid by working with the equations for the null horizon. The reason is that we parametrized the membrane equations in terms of the thermodynamic variables associated with the boundary fluid and required the metric $G_{\mu \nu}$ to match the boundary metric (in this case $\eta_{\mu\nu}$). In principle, one can also write the equations in terms of variables associated with a generic $r=const.$ hypersurface, thereby obtaining the hydrodynamics of those surfaces as described in \cite{RGflow}.

\end{document}